\documentclass[10pt]{iopart}

\usepackage{graphicx}
\expandafter\let\csname equation*\endcsname\relax
\expandafter\let\csname endequation*\endcsname\relax
\usepackage{amsmath}
\usepackage{algorithmic}
\usepackage{subfigure}
\usepackage{cite}
\usepackage{color}
\usepackage[normalem]{ulem}
\usepackage{dcolumn}% Align table columns on decimal point
\usepackage{bm}% bold math
\usepackage{bbm}
\usepackage{multirow}
\usepackage{url}

\begin{document}

\title[]{Efficient single pixel imaging in Fourier space}

\author{Liheng Bian, Jinli Suo, Xuemei Hu, Feng Chen and Qionghai Dai}

\address{Department of Automation, Tsinghua University, Beijing 100084, China}
\ead{jlsuo@tsinghua.edu.cn}
\vspace{10pt}
\begin{indented}
\item[]April 2016
\end{indented}

\begin{abstract}
Single pixel imaging (SPI) is a novel technique being able to capture 2D images using a bucket detector with high signal-to-noise ratio, wide spectrum range and low cost. Conventional SPI projects random illumination patterns to randomly and uniformly sample the entire scene's information. Determined by the Nyquist sampling theory, SPI needs either numerous projections or high computation cost to reconstruct the target scene, especially for high-resolution cases. To address this issue, {we propose an efficient single pixel imaging technique (eSPI)}, which instead projects sinusoidal patterns for importance sampling of the target scene's spatial spectrum in Fourier space. Specifically, utilizing the centrosymmetric conjugation and sparsity priors of natural images' spatial spectra, eSPI sequentially projects two $\frac{\pi}{2}$-phase-shifted sinusoidal patterns to obtain each Fourier coefficient in the most informative spatial frequency bands. eSPI can reduce requisite patterns by two orders of magnitude compared to conventional SPI, which helps a lot for fast and high-resolution SPI.
\end{abstract}

\vspace{2pc}
\noindent{\it Keywords}: single pixel imaging, computational ghost imaging, sinusoidal modulation, importance sampling
%\submitto{\JOPT}
%\maketitle
\ioptwocol

\section{Introduction}

Single pixel imaging (SPI) \cite{duarte2008single} is a novel {incoherent imaging} technique. {It produces 2D images using a bucket detector instead of array sensors. SPI shares the same imaging scheme with computational ghost imaging \cite{shapiro2008computational}, which uses a spatial light modulator (SLM) to generate programmable illumination patterns onto the target scene, and uses a bucket detector to collect the correlated lights.}
Then the 2D scene can be retrieved from the illumination patterns and corresponding 1D correlated single pixel measurements, using either linear correlation methods \cite{bromberg2009ghost, gong2010method, ferri2010differential, sun2012normalized} or compressive sensing (CS) techniques \cite{katz2009compressive, compressive_scientificreport}.
Due to its high signal-to-noise ratio, wide spectrum range, low cost and flexible light-path configuration, SPI has been widely applied in various fields \cite{zhao2012ghost, hu2015patch, 2016_SR_MSPI, gong2016three}.
%As reviewed by Erkmen and  Shapiro \cite{erkmen2010ghost}, the extension from conventional GI to SPI largely simplifies the physical implementation and broadens applications of ghost imaging systems.
%For conventional SPI, numerous programmable (usually by a spatial light modulator) illumination patterns are sequentially projected onto the scene, and the correlated intensities are collected by a single-pixel detector. Then the scene can be reconstructed using the patterns and their corresponding single-pixel measurements.

Despite the above advantages over conventional imaging techniques using array sensors, SPI needs numerous illumination patterns to reconstruct an image, which makes it time consuming and memory demanding \cite{erkmen2009signal}. Such a large number of patterns is caused by the utilized random modulation, which randomly and uniformly samples all {the target} scene's information with no discrimination. Determined by the Nyquist sampling theory, it needs at leat N measurements to reconstruct an N-pixel image. Especially, more measurements are needed in real applications to compensate for {the} system noise and {the} influences from other external factors. As a reference, Sun et al.\cite{sun20133d} used around $10^6$ patterns (20 times of {the} image pixels) to reconstruct a 256$\times$192-pixel image owning sufficient quality for subsequent 3D imaging. Though one can utilize compressive sensing \cite{katz2009compressive} to reduce projections, this largely increases computation complexity\cite{compressive_scientificreport}.
%In a nutshell, conventional SPI {needs} either numerous projections or high computation cost. This largely limits its practical applications. 
{Instead of using random patterns, the technique recently proposed in ref. \cite{Nature_2015} utilizes sinusoidal modulation to sample the scene's information in Fourier space. Specifically, it projects four $\frac{\pi}{2}$-phase-shifted patterns to sample each spatial frequency of the scene's spatial spectrum, and can save a lot of projections compared to conventional SPI.}

\begin{figure}[t]
  \centering
  \includegraphics[width=\linewidth]{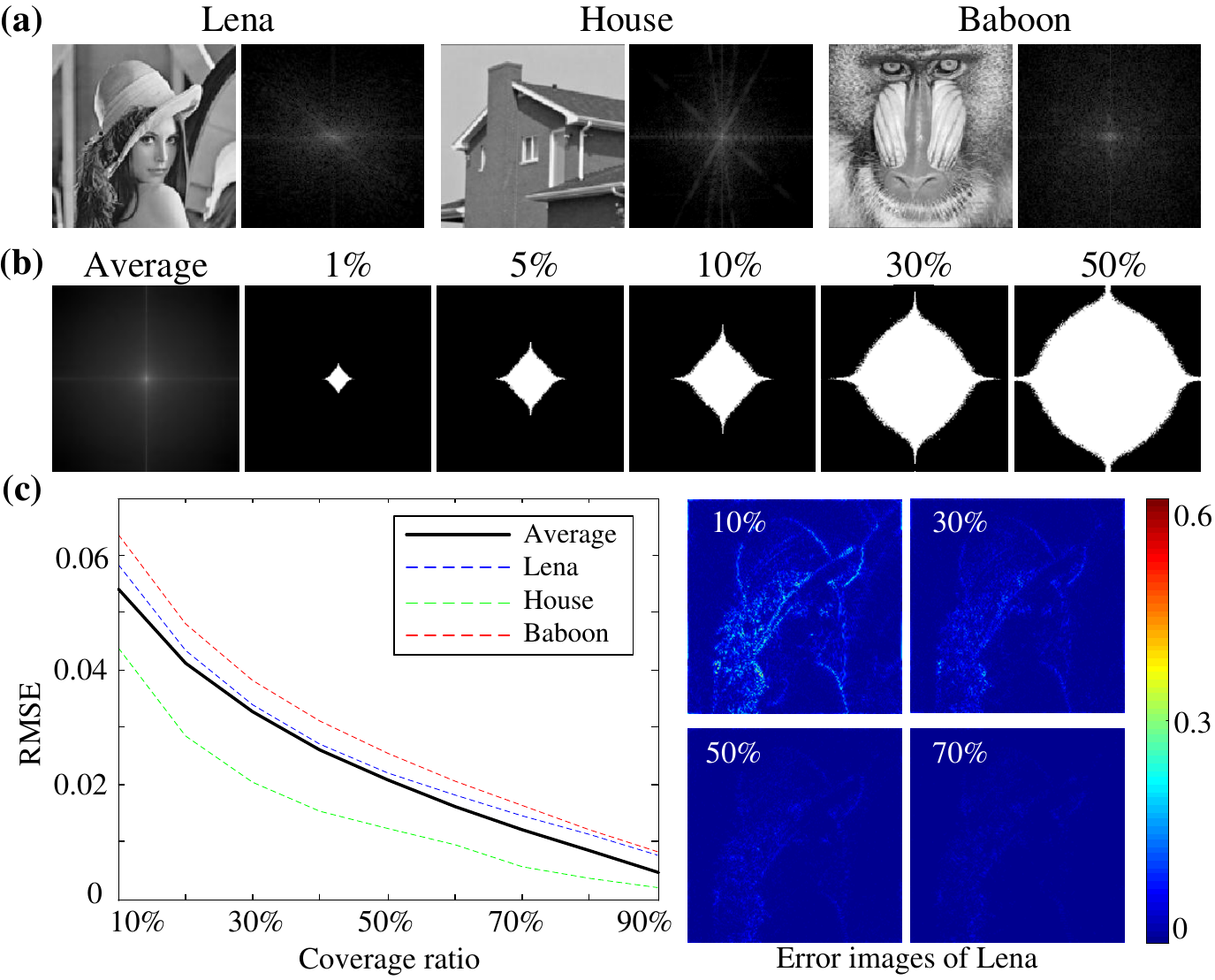}
  \caption{{Statistical study of natural images' spatial spectra.} (a) Three exemplar natural images and their spatial spectra. (b) The average spectrum of the USC-SIPI database, as well as different acquisition bands under different coverage ratios. (c) The relationship between reconstruction error and coverage ratio.
   }\label{fig:Prior}
\end{figure}

From the statistics\cite{Prior}, most information of natural images is concentrated in low spatial frequency bands and exhibits strong sparsity in Fourier space, as shown in Fig.~\ref{fig:Prior}(a) where several exemplar images and their spatial spectra are presented. {This motivates us to utilize the importance sampling strategy for efficient acquisition in Fourier space.} To realize the non-uniform sampling of the scene's spatial spectrum, we calculate the statistical importance distribution of nature images' spatial frequencies and sample them in a descending order of importance. To sample each spatial frequency, since random patterns do not work anymore, we use a two-step sinusoidal illumination modulation strategy similar to ref.\cite{khamoushi2015sinusoidal}, which is based on the centrosymmetric conjugation property of real natural images' spatial spectra. To conclude, {we propose an efficient single pixel imaging technique (eSPI) in this paper}. The technique utilizes the sparsity and conjugation priors of natural images' spatial spectra to realize fast SPI with extremely high efficiency and low computation cost.
{We note that the proposed eSPI differentiates from ref. \cite{Nature_2015} in two aspects: (i) utilizing the sparsity prior of natural images' spatial spectra, eSPI performs importance sampling in the Fourier domain, i.e., eSPI doesn't sample all the Fourier coefficients exhaustively as ref. \cite{Nature_2015}; (ii) incorporating the centrosymmetric conjugation property of natural images' spatial spectra into the patterning strategy, eSPI needs only two sinusoidal $\frac{\pi}{2}$-phase-shifted patterns for each frequency, instead of four as in ref. \cite{Nature_2015}. Benefitting from these two strategies, eSPI can save most projections of ref. \cite{Nature_2015}.}
%Technically, eSPI dose not project random patterns as conventional SPI. Instead, it designs sinusoidal illumination patterns for projection. Each two $\frac{\pi}{2}$-phase-shifted sinusoidal patterns owning the same single spatial frequency are sequentially projected onto the target scene, which yield the target image's Fourier coefficient at the same frequency.
%By sequentially projecting these sinusoidal pattern groups owning spatial frequencies in the statistically most informative spectrum band, the target image's spatial spectrum can be obtained. Then inverse Fourier transform is applied to reconstruct the scene.
In the following, we begin to introduce eSPI in two steps.

\section{Methods}

The first step of eSPI is to determine {the} acquisition band {in Fourier space}, i.e., to decide which Fourier coefficients to sample.
Here we first study {the} statistical distribution of natural scenes' {spatial} spectra, and accordingly determine the priority of spectrum sampling. Specifically, we transform all the 44 images in {the} USC-SIPI common miscellaneous database \cite{Dataset_1} to Fourier space, and calculate the spectra's average magnitude map, as shown in the first image in Fig.~\ref{fig:Prior}(b). Then we threshold it to determine the acquisition bands under different coverage ratios (the ratio between {the} acquisition band and {the} whole spectrum). The results are shown in {Fig.~\ref{fig:Prior}(b)}, where {the} white areas stand for {the} acquisition bands.
%Then we do binary thresholding to the map by setting small values to 0 and large values to 1 (we call the areas of 0 inactive frequencies, and the areas of 1 active frequencies), with different thresholds. Thus, we get the general frequency sampling maps under different ratios of active frequencies in the Fourier domain, as shown in Fig. \ref{fig:Prior_Distribution}.

Based on the thresholding results, users can determine {the} acquisition band by setting different coverage ratios according to specific applications. Larger coverage ratio results in a wider acquisition band and more detailed information, but more projections.
To further study the relationship between coverage ratio and reconstruction error, we successively sample the {spatial} spectrum of each image in the above dataset under different coverage ratios, transform them back to spatial space, and calculate reconstruction errors in terms of root-mean-square error (RMSE). {RMSE is defined as $\sqrt{E(({\bf I}_1-{\bf I}_2)^2)}$ to measure the difference between two images ${\bf I}_1$ and ${\bf I}_2$, where $E$ is the pixel-wise average operation.} The average performance is plotted as the black solid curve in Fig. \ref{fig:Prior}(c), where reconstruction errors of several exemplar images are also plotted with dashed lines. The results indicate that though different images are of slight diversity, they follow the same trend {that reconstruction error decreases as coverage ratio increases}. Besides, the reconstruction residues of {the} ``Lena" {image} at different coverage ratios are also presented as a reference.
%In the rest of the paper, we preserve top 30$\%$ frequencies in the proposed eSPI as an example.

%\textbf{Sample the coefficients.}
%With the active frequencies determined, in the next, we design a pattern projecting strategy to sequentially sample their coefficients.
%Following the Fourier transform definition, a pattern conducting delta sampling in the Fourier domain corresponds to a 2D sine patterns with a specific frequency and direction. Then, utilizing the sparsity of nature images in the Fourier domain, the proposed eSPI can largely reduce the requisite number of projected patterns.
%In the following, we will explain the principle of eSPI, both synthetic and real experiments, and extended discussion in details.
%frequency of each composite (i.e., random pattern) in conventional single pixel imaging and the result is illustrated in Fig.~\ref{fig:random}. Intuitively, each binary pattern encodes a random combination of frequencies of the latent sharp image. The following mathematical derivation also consists with above illustration.
%\begin{equation}
%\mathcal F ({\bf I} \odot {\bf p} )=\mathcal F({\bf I}) \odot \mathcal F({\bf p}),
%\end{equation}
%in which $\bf I$ and $\bf p$ respectively represent an image and a pattern of the same size as $\bf I$, $\odot$ denotes element-wise product, and $\mathcal F$ denotes Fourier transform.

After the acquisition band determined, we move on to the second step of eSPI, i.e., { sampling each Fourier coefficient in the band to perform the non-uniform acquisition. Since random patterns do not work anymore, we use a two-step sinusoidal illumination modulation strategy similar to ref. \cite{khamoushi2015sinusoidal} based on the centrosymmetric conjugation property of real natural images' spatial spectra. }
To introduce the illumination patterning strategy in detail, we first analyze the information encoded by the single pixel measurements in Fourier space.
According to the Fourier theorem, a 2D image $\bf I$ can be represented as ${\bf I} = \sum_{i}c_i{\bf B_i}$, where $\bf B_i$ is the $i$th normalized Fourier basis, and $c_i$ is its Fourier coefficient.
%To further illustrate the Fourier presentation and Fourier patterns, we present several examples in Fig.~\ref{fig:FT_Example}. In Fig.~\ref{fig:FT_Example}(a), we use the Fourier basis images (correspond to $\bf B_i$) to facilitate better understanding of the Fourier presentation, with several Fourier basis examples standing for different spatial frequencies and phases are shown in Fig.~\ref{fig:FT_Example}(b).
Similarly, by applying Fourier transform to a projected pattern $\bf P$, we can get ${\bf P} = \sum_{j}\hat{c}_j{\bf B_j}$. Its corresponding single pixel measurement $s$ can be represented as
\begin{eqnarray}\label{eqs:Measurement_1}
%s &=& |\sum\sum{\bf I}\odot{\bf P}^*|\\\nonumber
s &=& |\sum_{m}\sum_{n}{\bf I}(m,n){\bf P}(m,n)|\\\nonumber
&=& |\sum_{m}\sum_{n}[\sum_{i}c_i{\bf B}_i(m,n)][\sum_{j}\hat c_j{\bf B}_j(m,n)]|\\\nonumber
&=& |\sum_{i}\sum_{j}c_i \hat c_j[\sum_{m}\sum_{n}{\bf B}_i(m,n){\bf B}_j(m,n)]|.\\\nonumber
\end{eqnarray}
Here $(m,n)$ index the 2D spatial coordinate.
Substituting the orthogonality of the Fourier bases
\begin{eqnarray}
f(x)=\left\{
\begin{aligned}
\sum_{m}\sum_{n}{\bf B}_i(m,n){\bf B}_j(m,n) & =0, & i\neq j\\
\sum_{m}\sum_{n}{\bf B}_i(m,n){\bf B}_j(m,n) & =1, & i = j,
\end{aligned}
\right.
\end{eqnarray}
into the above equation, we get
\begin{eqnarray}\label{eqs:Measurement_2}
s &=& |\sum_{j}c_j \hat c_j|.
\end{eqnarray}
From this we can see that $\{\hat c_j\}$ is a spectrum sampling vector to record the scene's Fourier coefficients.
%$s$ is the summation of the scene's frequency coefficients weighted by the conjugate coefficients of the projected pattern's frequencies.
Therefore, we can directly sample a specific Fourier coefficient by setting $\{\hat c_j\}$ as a delta vector (containing only one non-zero entry), which results in a sinusoidal pattern with complex values.
%Thus if we project a pattern corresponding to a standard single frequency in the Fourier domain, i.e., only one entry of \{$c'$\} is not equal to 0, then the measurement of the single-pixel camera is exactly the coefficient of the same spatial frequency of the scene. This is the first foundation of the proposed eSPI. In the following we call this kind of patterns as Fourier patterns to differentiate them from the conventional random patterns.
%{\em Using Fourier patterns, the computation ghost imaging algorithms consists with the principal of Fourier transform.}

\begin{figure}[!t]
  \centering
  \includegraphics[width=0.6\linewidth]{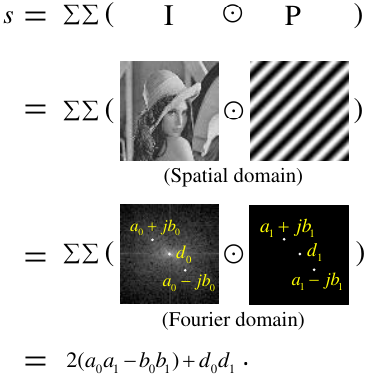}
  \caption{Illustration of the encoded information in a correlated single pixel measurement when a real valued sinusoidal pattern is projected.}\label{fig:Sampling_Example}
\end{figure}

However, real facilities can only project real-valued sinusoidal patterns, each owning three non-zero coefficients in its spatial spectrum---two conjugate coefficients of a centrosymmetric non-zero frequency pair and one of the zero frequency. The conjugation property also holds for natural scenes. Let $c_1 = a_0 + jb_0$, $c_2 = a_0 - jb_0$ and $c_3 = d_0$ {($j$ is the imaginary unit)} denote the three non-zero coefficients of the target scene $\bf I$, and $\hat c_1 = a_1 + jb_1$, $\hat c_2 = a_1 - jb_1$ and $\hat c_3 = d_1$ represent corresponding coefficients of a sinusoidal pattern $\bf P$, we have
\begin{eqnarray}
s \!&=&\!|c_1 \hat c_1 + c_2 \hat c_2+c_3 \hat c_3|\\\nonumber
\!&=&\! |(a_0 \!+\! jb_0)(a_1 \!+\! jb_1) + (a_0 \!-\! jb_0)(a_1 \!-\! jb_1)+d_0d_1|\\\nonumber
\!&=&\! 2(a_0a_1 - b_0b_1)+d_0d_1.
\end{eqnarray}
A more explicit demonstration is shown in Fig. \ref{fig:Sampling_Example}. {Note that if the pattern's pixel number in each dimension is even, determined by the symmetry property of discrete Fourier transform, there is no corresponding centrosymmetric counterpart of the highest spatial frequency, i.e., the highest frequency cannot form a conjugation frequency pair.}

Based on the above derivations, acquiring a specific Fourier coefficient turns into computing $a_0$ and $b_0$, with $s, a_1, b_1$ and $d_1$ known. To achieve this, we sequentially project three patterns onto the target scene. The first one is a uniform pattern with the constant intensity equal to the mean pixel value of $\bf P$, and the measurement is exactly $d_0d_1$.
The other two patterns are sinusoidal patterns with Fourier coefficients being $\{a_1 = \frac{1}{2}, b_1 = 0, d_1 = 1\}$ and $\{a_1 = 0, b_1 = \frac{1}{2}, d_1 = 1\}$, respectively. Thus we can obtain $a_0$ and $b_0$ by simply subtracting $d_0d_1$ from the correlated measurements. %, and thus the two coefficients of the scene's spectrum corresponding to the specific frequency.

\begin{figure*}[!t]
  \centering
  \includegraphics[width=0.8\linewidth]{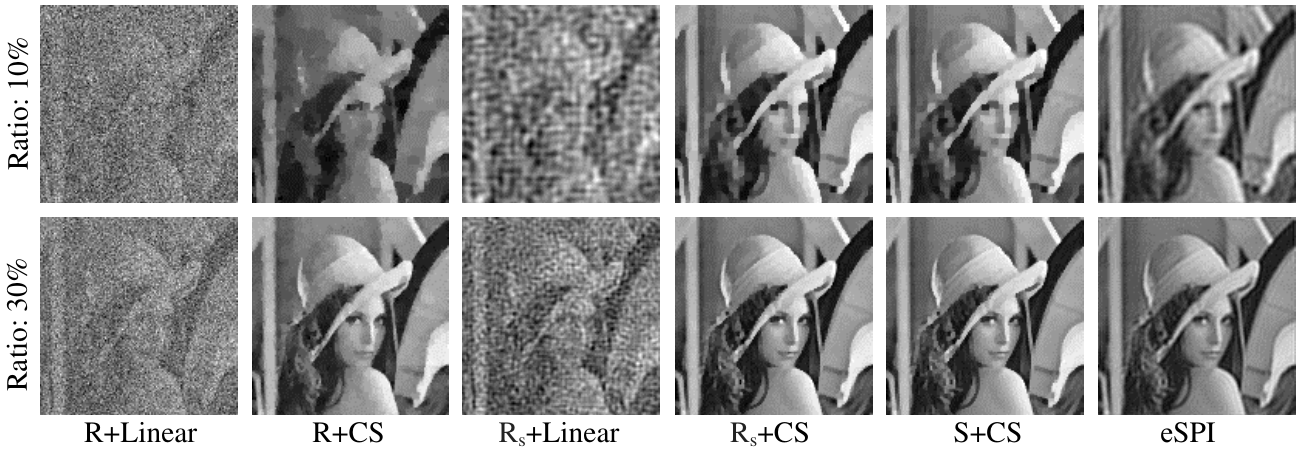}\\
  \caption{Simulated reconstruction results of the Lena image (128$\times128$ pixels) by different SPI strategies, with the coverage ratio being 10$\%$ and 30$\%$, respectively. ``R", ``R$_s$", ``S", ``Linear'' and ``CS'' stand for random modulation, {random modulation of the same speckle traverse size (same spatial frequency) as eSPI}, sinusoidal modulation, linear correlation reconstruction, and compressive sensing reconstruction, respectively.}\label{fig:Syn_Result}
\end{figure*}

Following the above method, we can obtain all the Fourier coefficients of the pre-determined acquisition band, by sequentially projecting corresponding sinusoidal patterns (the uniform pattern needs to be projected only once for all the frequencies). Then, the target scene can be recovered by inverse Fourier transform to the obtained spatial spectrum.

\section{Results}

To validate the proposed eSPI technique, we first conduct a simulation experiment to compare {the} reconstruction performance of different SPI methods.
We set the ``Lena" image ($128\times128$ and $256\times256$ pixels respectively) as the latent target scene image, and synthesize {the} measurements of different patterns following Eq. \ref{eqs:Measurement_1}. We set the coverage ratio {being} 0.1 and 0.3 (corresponding acquisition bands are shown in the {fourth and fifth} subfigures in Fig. \ref{fig:Prior}(b)), {respectively}. The experiment is conducted using Matlab on an Intel i7 3.6GHz CPU computer, with 16G RAM and 64 bit Windows 7 system. 
%For comparison, we respectively use the {differential ghost imaging method (DGI)} \cite{ferri2010differential} and the augmented Lagrange multiplier algorithm (ALM) \cite{ALM} to represent {the} linear correlation methods and {the} compressive sensing {(CS)} methods considering their satisfying reconstruction performance. 
For comparison, the linear correlation based reconstruction method \cite{gong2010method, ferri2010differential} and the compressive sensing based technique \cite{duarte2008single} are applied on the same set of sinusoidal patterns, as well as the same number of random patterns. {Also, we compare eSPI with conventional SPI in the sense of the same speckle transverse size \cite{gong2016three} (same spatial frequency), by truncating conventional random patterns' spatial spectra with the same acquisition band (Fig. \ref{fig:Prior}(b)) as eSPI.} The results are shown in Fig. \ref{fig:Syn_Result} and Tab. \ref{tab:Syn_Comparision}. Note that we omit the results of ``S+Linear", since {the} eSPI reconstruction (namely inverse Fourier transform) is essentially a linear combination of {the} Fourier bases, which is intrinsically the same as the linear correlation based method in {the case} of sinusoidal patterns.

\begin{table}[t]\footnotesize
  \centering
  \caption{Quantitative comparison among different SPI strategies under different coverage ratios and image sizes. The "$\times$" symbol means that the reconstruction is out of memory.}
  \begin{tabular}{cllclc} \\ \hline
    &  & \multicolumn{2}{c}{Ratio: 10$\%$} & \multicolumn{2}{c}{Ratio: 30$\%$} \\
   &  & RMSE & Time & RMSE & Time \\ \hline
  \multirow{4}{3em}{128$\times$128\\pixels} & R+Linear & 0.215 & 2s & 0.191 & 6s \\
   & R+CS & 0.115 & 68min & 0.042 & 92min \\
   & R$_s$+Linear & 0.203 & 2s & 0.187 & 6s \\
   & R$_s$+CS & 0.075 & 68min & 0.041 & 91min \\
   & S+CS & 0.066 & 67min & \textbf{0.037} & 92min \\
   & eSPI & \textbf{0.061} & \textbf{1s} & 0.044 & \textbf{3s} \\\hline
   \multirow{4}{3em}{256$\times$256\\pixels} & R+Linear & 0.211 & 9s & 0.188 & 26s \\
   & R+CS & $\times$ & $\times$ & $\times$ & $\times$ \\
   & R$_s$+Linear & 0.205 & 9s & 0.186 & 25s \\
   & R$_s$+CS & $\times$ & $\times$ & $\times$ & $\times$ \\
   & S+CS & $\times$ & $\times$ & $\times$ & $\times$ \\
   & eSPI & \textbf{0.035} & \textbf{3s} & \textbf{0.014} & \textbf{8s} \\\hline
  \end{tabular}\label{tab:Syn_Comparision}
\end{table}

From both the visual and quantitative results, we can clearly see that eSPI largely outperforms conventional SPI in terms of both efficiency and reconstruction quality. {The advantages come from the utilized sparse information encoding strategy. For conventional SPI, the spatial spectra of random patterns are also random. They sample and multiplex the target scene's whole spectrum randomly and uniformly with no discrimination. Thus conventional SPI can not utilize the importance sampling strategy, and need much more projections for demulplexing and reconstruction. Instead, each sinusoidal pattern in eSPI only encodes a Fourier coefficient pair of the scene's spatial spectrum. Based on this, eSPI samples only the most informative bands and omits unimportant ones. Therefore, it is much more efficient.}
Note that though the compressive sensing (CS) method produces similar results as eSPI when using sinusoidal patterns, it is much more time consuming and memory demanding. Especially, when image size grows large enough, CS does not work anymore. This is because CS models the reconstruction as an ill-posed problem, which needs large memory and long time for computation under an optimization framework. Instead, eSPI is linear correlation based and doesn't involve any complex calculations, so it is much faster and memory saving.

\begin{figure}[h]
  \centering
  \includegraphics[width=0.9\linewidth]{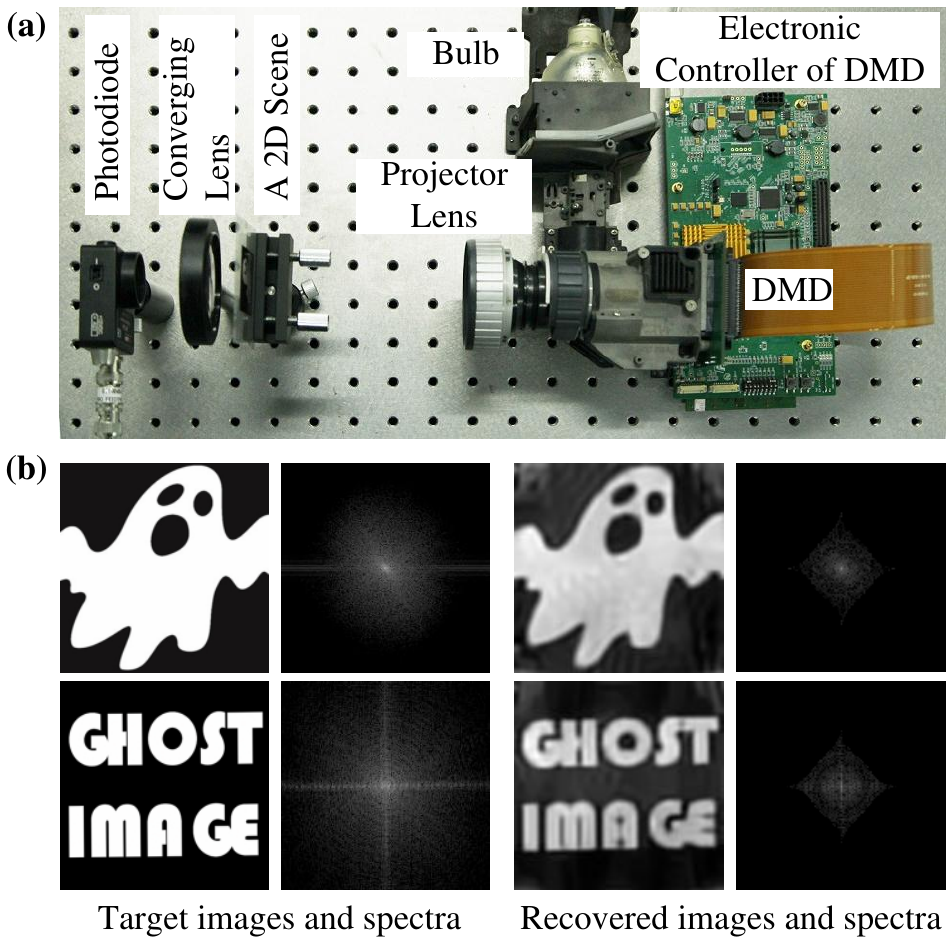}
  \caption{Experiment on real captured data. (a) The eSPI prototype. (b) Reconstruction results of two different scenes (each owning128$\times$128 pixels) with the coverage ratio being 10$\%$. {The left two columns are the ground-truth target images and their spatial spectra, and the right two columns are corresponding reconstruction.}}\label{fig:Real}
\end{figure}

To further validate eSPI, we build a {proof-of-concept} prototype exhibited in Fig.~\ref{fig:Real}(a). The system mainly consists of two parts including programmable illumination and detection. The illumination part includes a commercial projector's illumination module (numerical aperture of the projector lens is 0.27) and a digital micromirror device (DMD, Texas Instrument DLP Discovery 4100 Development Kit, .7XGA) for spatial modulation. We use the 8-bit mode of the DMD to generate patterns, with the frame rate being 30Hz.
Patterns owning 128$\times$128 pixels are sequentially projected onto a printed transmissive film (34mm$\times$34mm) as the target scene.
Then the correlated lights are recorded by a high-speed bucket detector (Thorlabs DET100 Silicon photodiode, 340-1100 nm) with a 14-bit acquisition board ART PCI8514. The sampling rate is set as 10kHz. We utilize the self-synchronization technique in ref. \cite{suo2015self} to synchronize the DMD and the detector. For each pattern, we average all its corresponding stable measurements for subsequent reconstruction. The coverage ratio of the acquisition band is set as $10\%$, resulting in 1635 projected patterns in total.
The reconstructed results of two different scenes are shown in Fig.~\ref{fig:Real}(b), from which we can see that 10$\%$ of the pixel number patterns can yield satisfying results. Compared to ref.\cite{sun20133d} {where the requisite pattern number is 20 times of {the} pixel number}, eSPI can reduce projections by two orders of magnitude. {Note that there exist some artifacts in the reconstructed images. This may be caused by several factors, including film glare, light flicker (voltage fluctuation), ambient light, modulation deviation of the DMD, thermal noise of the detector, and so on. Further efforts are needed to address these problems by improving the experimental environment and imaging elements, and proposing noise-robust reconstruction techniques.}

\section{Conclusion and discussion}

In this paper, {we propose an efficient single pixel imaging technique (eSPI)}. {Different from conventional random illumination modulation which randomly and uniformly samples the scene's whole spatial spectrum, eSPI uses a two-step sinusoidal illumination modulation strategy to obtain the Fourier coefficients of the target scene's most informative spectrum band.} As a result, we can reduce the requisite patterns by two orders of magnitude. This helps a lot for fast and high resolution SPI.

\begin{figure}[!t]
  \centering
  \includegraphics[width=\linewidth]{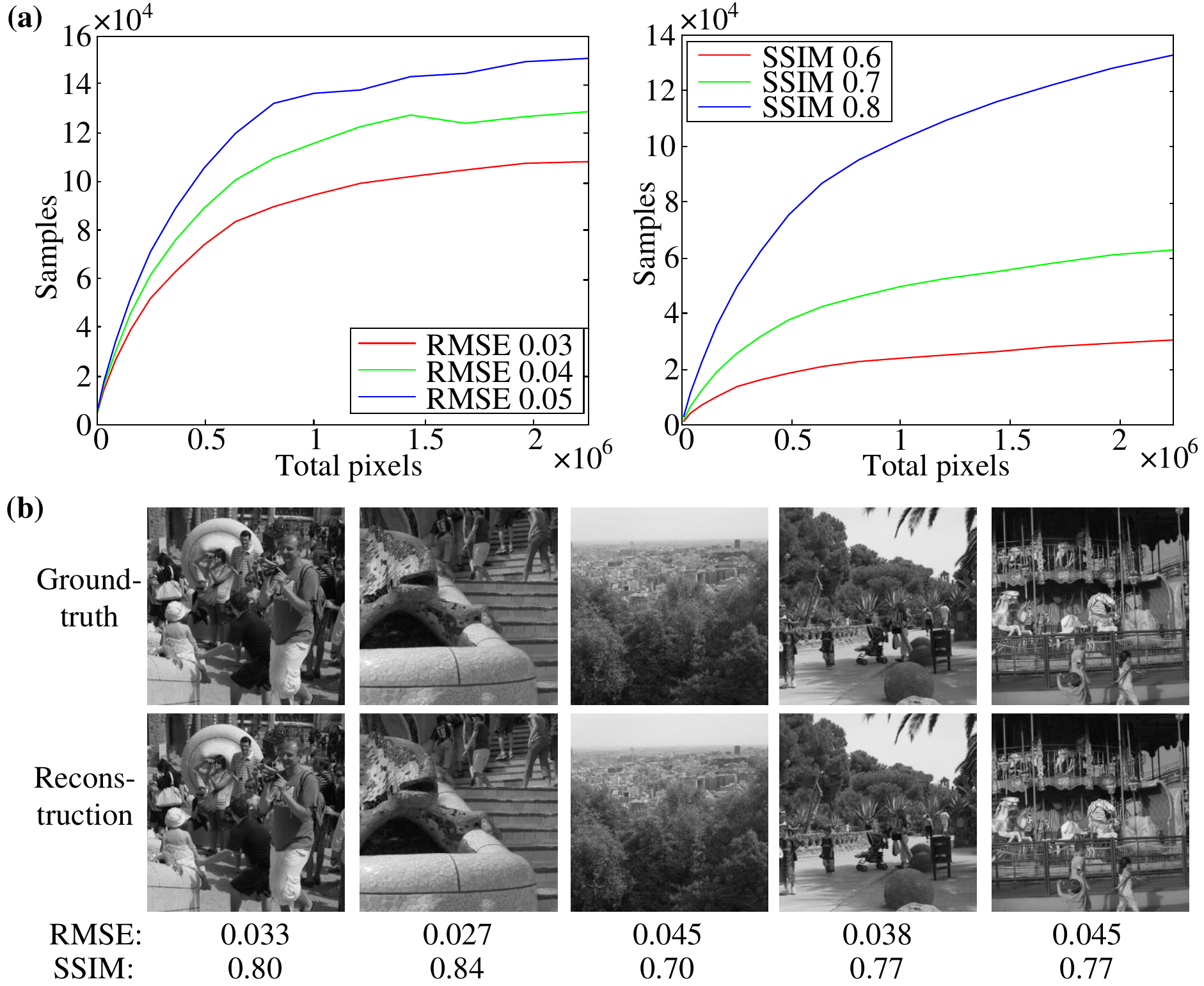}
  \caption{{Demonstration of eSPI's advantages for high resolution imaging.} (a) Required samplings at different image sizes for the same reconstruction quality. (d) Exemplar reconstructed megapixel images using $10^5$ samplings.
   }\label{fig:Stat}
\end{figure}

Due to the utilized importance sampling strategy, eSPI owns more advantages when applied to high resolution imaging, where the images' spatial spectra are more sparse. To demonstrate this, we downsample each of the 322 natural images (2268$\times$1512 pixels) in the Barcelona Calibrated Images Database \cite{Dataset_Barcelona} to different image sizes, and successively sample their spatial spectra under different coverage ratios. Then we transform them back to spatial space, and quantify the reconstruction quality in terms of RMSE and {the} structure similarity index (SSIM) \cite{SSIM}. SSIM measures {the} structural similarity between two images. It ranges from 0 to 1, with larger amount meaning more similar structure. %The two metrics respectively measure reconstruction error and similarity.
As shown in Fig. \ref{fig:Stat}(a), the required sampling number increases slower for the same reconstruction quality as the image size grows. This means that for high-resolution imaging, linearly increased samplings are unnecessary. Specifically, around $10^5$ samplings are enough to retrieve a megapixel image with satisfying visual quality, as shown in Fig. \ref{fig:Stat}(b).
{We want to note that the low sampling frequencies of eSPI are not caused by the hardware limit. Instead, it is determined by the utilized importance sampling strategy for much higher efficiency with no degeneration of final reconstruction.}

eSPI can be widely extended. Since the measurement formation in Eq. (\ref{eqs:Measurement_1}) is linear, we can adopt multiplexing \cite{multiplexing} to raise {the} signal-to-noise ratio of final reconstruction. Besides, the content-adaptive sampling scheme \cite{bian2014content} can be introduced for higher efficiency. In addition, there exist many other generative image representation methods such as {the} discrete cosine transform. It is interesting to study the pros and cons by applying these transforms to the proposed eSPI framework. What's more, as the requisite number of illumination patterns is largely reduced, eSPI offers promising potentials for real time SPI. These are our future work.

%In summary, this letter exploits the sparsity and conjugation properties of natural images' spatial spectra, and proposes a novel SPI strategy termed eSPI, which sequentially projects two $\frac{\pi}{2}$-phase-shifted sinusoidal patterns for each centrosymmetric frequency pair in Fourier space, to conduct importance sampling of a scene's spatial spectrum. eSPI can reduce the requisite number of projected patterns by two orders of magnitude compared to conventional SPI. Both the simulation and real experiments validate the effectiveness of eSPI.

\section*{Acknowledgements}

This work was supported by the National Natural Science Foundation of China (Nos. 61327902 and 61120106003).


\section*{References}

\bibliographystyle{unsrt}
\bibliography{FSPI}

\end{document}